\documentclass[grg,amsmath,amssymb]{revtex4}
\usepackage[dvips]{graphicx}
\usepackage{dcolumn}% Align table columns on decimal point
\usepackage{bm}% bold math

\begin{document}

\title[The dragging of inertial frame in the HYPER
project]{Gravitational perturbations on local experiments in a
  satellite~: \\ The dragging of inertial frame in the HYPER project.}

\author{M-C. Angonin-Willaime}
\email{m-c.angonin@obspm.fr}
\author{X. Ovido}
\author{Ph. Tourrenc}
\affiliation{
Universit\'{e} P. et M. Curie\\
ERGA, case 142\\
4, place Jussieu\\
F-75252 Paris CEDEX 05, France}

\date{\today}

\begin{abstract}
We consider a nearly free falling Earth satellite where atomic wave
interferometers are tied to a telescope pointing towards a faraway star.
They measure the acceleration and the rotation relatively to the local
inertial frame.

We calculate the rotation of the telescope due to the aberrations and the
deflection of the light in the gravitational field of the Earth. We show
that the deflection due to the quadrupolar momentum of the gravity is not
negligible if one wants to observe the Lense-Thirring effect of the Earth.

We consider some perturbation to the ideal device and we discuss the orders
of magnitude of\ the phase shifts due to the residual tidal gravitational
field in the satellite and we exhibit the terms which must be taken into
account to calculate and interpret the full signal.

Within the framework of a geometric model, we calculate the various periodic
components of the signal which must be analyzed to detect the Lense-Tirring
effect. We discuss the results which support a reasonable optimism.

As a conclusion we put forward the necessity of a more complete, realistic
and powerful model in order to obtain a final conclusion on the theoretical
feasibility of the experiment as far as the observation of the
Lense-Thirring effect is involved.

\end{abstract}
\pacs{04.25.Nx, 04.80.Cc, 07.60.Ly, 95.30.Sf}

\maketitle

\section{Introduction}

The quick development of atomic interferometry during the last ten
years is impressive. The clocks, the accelerometers and the gyroscopes
based on this principle are already among the best that have been
constructed until now and further improvements are still expected.
This situation favors a renewal in the conception of various
experiments, such as the measurement of the fine structure constant or
the tests of relativistic theories of gravitation currently developed
by classical means (gravitational frequency shifts, equivalence
principle\footnote{ MICROSCOPE \cite{Touboul} is a CNES mission
  designed to compare the motion of two free falling macroscopic
  masses in order to check the equivalence principle. It has been
  decided and should be launched in not too far a future (except for
  any possible delay!). Several other ''classical'' and more ambitious
  projects are also considered \textit{i.e.} STEP (for more details,
  see e.g. : einstein.stanford.edu/STEP/index.html) and Galileo
  Galilei \cite{Nobili}}, Lense-Thirring effect\footnote{
  Lense-Thirring effect originates in the diurnal rotation of the
  Earth. It results in an angular velocity that a gyroscope, pointing
  towards a far away star, can measure. Lense-Thirring angular
  velocity depends on the position of the satellite. GPB ( Gravity
  Probe B; for more details, see e.g. : einstein.stanford.edu) is a
  NASA project designed to measure the secular precession of a
  mechanical gyroscope due to Lense-Thirring effect. It has been
  carefully studied for many years at Stanford University, it is now
  expected to be launched in a near future.}, etc.$\ldots $).

The performances of laser cooled atomic devices is limited on Earth by the
gravity. Therefore further improvements demand that new experiments take
place in free falling (or nearly free falling) satellites. A laser cooled
atomic clock, named PHARAO, will be a part of ACES (Atomic Clock Ensemble in
Space), an ESA mission on the ISS planned for 2006. Various other
experimental possibilities involving ''Hyper-precision cold atom
interferometry in space'' are presently considered. They might result in a
project (called ''Hyper'') in not too far a future.
Most of the modern experiments display such a high sensitivity that their
description must involve relativistic gravitation. This is not only true for
the experiments which are designed to study the gravitation itself but also
for any experiment such as Hyper where very small perturbations cannot be
neglected any longer.

The present paper is a contribution to the current discussions on the
feasibility of Hyper. We consider especially the\ effect of the inertial
fields and the local gravitational fields in a satellite\footnote{
Both fields are called ''gravitational fields'' in the sequel.}.

There are two kinds of gravitational perturbation.

\begin{enumerate}
\item The masses in the satellite produce a gravitational field which is not
negligible. In some experiments, the mass distribution itself can play a
role : This is, for instance, the case for GPB. However, some other
experiments are only sensitive to the change of the mass distribution with
the time. This is the case of Hyper where a signal is recorded as a function
of the time and analyzed by Fourier methods at a given frequency. The
modification of the mass distribution is due to mechanical and thermal
effects. It depends on the construction of the satellite, the damping of the
vibrations and the stabilization of the temperature. We will not study these
effects which can be considered as technological perturbations. We do not
claim that these perturbations are easy to cancel but only that it is
possible in principle while it is impossible for tidal effects from the
Earth.

\item The perturbations due to the gravity of far away bodies (the Earth,
the Moon, the Sun and the surrounding planets) is the subject of the present
paper. It is impossible to cancel their action.
\end{enumerate}

The aim of this paper is to study the gravity in a nearly free falling
satellite where the tidal effects remain.

The experimental set-up is tied to a telescope pointing towards a ''fixed
star''. However, it experiences a rotation: the so called ''Lense-Thirring''
effect. It has been recently noticed that atomic interferometers display a
sensitivity high enough to map the gravitomagnetic field of the Earth
(included the Lense-Thirring effect). This could be one of the goals of the
Hyper project \cite{Raselhyper}. The effect is so tiny that we will concentrate on this
question.

We consider the case where the experimental set up is built out of several
atomic interferometers similar to those which are currently developed in
Hannover \cite{Rasel1} and \cite{Rasel2}, Paris \cite{belarno} and Orsay \cite{Bouyer1} and \cite{Bouyer2} (see section \ref{asu}).

\ 

In section 1 we introduce the metric, $g_{\alpha \beta },$ in the non
rotating geocentric coordinates and we define a book-keeping of the orders
of magnitude.

In order to study the local gravitational field in the satellite, we chose
an origin, $O,$ and a tetrad $u_{\left( \sigma \right) }^{\alpha }$ which
defines the reference frame of the observer at point $O.$ The time vector, $%
u_{\left( 0\right) }^{\alpha },$ is the 4-velocity of $O.$ The space vector $%
u_{\left( 1\right) }^{\alpha }$ defines the axis of a telescope which points
towards a ''fixed'' far away star. The tetrad is spinning around $u_{\left(
1\right) }^{\alpha }$ with the angular velocity, $\varpi .$

In section 2, in order to define precisely the tetrad we study the apparent
direction of the star.

Then, in section 3, following \cite{LiZ} and \cite{LiNi} we expand the
metric in the neighborhood of $O$ (the NiZiLi metric).

Finally we calculate the response of the experimental set-up and we
emphasize the interest of spinning the satellite.

An ASU delivers a phase difference, $\delta \varphi ,$ between two matter
waves. The phase difference is the amount of various terms. Some of them can
be computed with the required accuracy; they produce a phase difference $%
\delta \varphi _{k}$. Then $\delta \varphi =\delta \varphi _{k}+\delta
\varphi _{u}$ where $\delta \varphi $ is measured. Therefore one can
consider that the ASU delivers $\delta \varphi _{u}.$ This is this quantity
that we want to calculate here.

In this paper, we point out the various contributions to $\delta \varphi
_{u} $ with their order of magnitude. The method that we use to calculate $%
\delta \varphi _{u}$ is a first order perturbation method$.$ A more precise
method, valid for $\delta \varphi $ is now available \cite{Antoine}. It gives the
possibility to model the ASU and therefore to study the signal due to the
various perturbations which are expected.

\section{Generalities}

\subsection{Conventions and notations}

In the non rotating geocentric frame we introduce the coordinates $x^{\alpha
}$ with $\alpha =0,1,2,3.$ We define the time coordinate $x^{0}=ct$ where $c$
is the celerity of the light in the vacuum. The space coordinates are $x^{k}$
(in this paper the Latin indices run from $1$ to $3).$ We use the notations $%
\vec{r}=\left\{ x^{k}\right\} =\left\{ x,y,z\right\} $ and we define the
spherical coordinates $\left\{ r,\theta ,\varphi \right\} ,$ \textit{i.e. }$%
x=r\sin \theta \,\cos \varphi \;,\;\;y=r\sin \theta \,\sin \varphi
\;,\;\;z=r\cos \theta .$ Therefore $r=\left\| \overrightarrow{r}\right\| =%
\sqrt{x^{2}+y^{2}+z^{2}}.$

We consider an Earth satellite and a point $O$ which is chosen as the origin
of the local coordinates in the satellite. We assume that the position of $O$
is given by its three space coordinates, $\vec{r}=\left\{ x,y,z\right\}
=\left\{ x^{k}\right\} ,$ as three known functions of the coordinate time, $%
t $. Then we define the velocity of point $O$ as $\overrightarrow{v}=\dfrac{d%
\vec{r}}{dt}.$

The proper time at point $O$ is $s.$ The motion of $O$ can be described as
well by the four functions $x^{\alpha }=x^{\alpha }\left( s\right) .$ The
four-velocity is defined as $u^{\alpha }=\dfrac{dx^{\alpha }}{ds}.$

The Newtonian constant of gravitation is $G.$

We use geometrical units where the numerical value of $G$ and $c$ is equal
to 1.

The metric tensor is $g_{\alpha \beta }$; it inverse is $g^{\alpha \beta }.$
The Minkowski tensor is $\eta _{\alpha \beta }=diag\left[ 1,-1,-1,-1\right] $%
; its inverse is $\eta ^{\alpha \beta }.$

We use the summation rule on repeated indices (one up and one down).

The partial derivative of $\left( \;\right) $ will be noted $\left(
\;\right) _{,\,\alpha }=\partial _{\alpha }\left( \;\right) =\dfrac{\partial
\left( \;\right) }{\partial x^{\alpha }}.$

In the sequel we introduce different tetrads, \textit{i.e.} a set of four
vectors $e_{\hat{\sigma}}^{\alpha}$  labeled by the means of an index called $\hat{\sigma}$ or $\left(
\sigma \right) ,$ such as $e_{\hat{\sigma}}^{\alpha }g_{\alpha \beta }e_{%
\hat{\rho}}^{\beta }=\eta _{\hat{\sigma}\hat{\rho}}$. The ''Einstein''
indices, $\alpha ,$ $\beta ,$ $\sigma ,$ etc., and the ''Minkowski'' indices 
$\hat{\alpha},$ $\hat{\beta},$ etc., can be changed one into the other by
the means of the tetrad : $\left( \;\right) _{\hat{\alpha}}=e_{\hat{\alpha}%
}^{\mu }\,\left( \;\right) _{\mu }.$ The metric tensor is used to raise (or
lower) the Einstein indices while the Minkowski tensor is used for Minkowski
indices.

\subsection{The fundamental element}

In the sequel we consider the \textsc{P}arametrized \textsc{P}ost \textsc{N}ewtonian theories \cite{will}. The relevant \textsc{PPN} parameters which
appear below are $\gamma $ and $\alpha _{1}.$ The parameter $\gamma $ is the
usual parameter connected to the deflection of a light ray by a central
mass. The parameter $\alpha _{1}$ couples the metric to the speed, $-\vec{w}%
, $ of the preferred frame (if any) relatively to the geocentric frame. In
general relativity, $\alpha _{1}=0$ and $\gamma =1.$

The preferred frame is generally considered to be the rest frame of the
Universe where the background radiation is isotropic ($\left\| \vec{w}%
\right\| \sim 10^{-3}$ in geometrical units).

Let us define now several quantities which will be used in the sequel~:

$\bullet \qquad 2M_{\oplus }$ is the Schwarzschild's radius of the Earth ($%
M_{\oplus }\simeq 4.4 \, {\rm mm}).$ As we use geometrical units (G = c = 1), $M_{\oplus }$
is also called the ''mass'' of the Earth.

$\bullet \qquad \vec{J}_{\oplus }$ is the angular momentum of the Earth in
geometrical units$.$ The relevant quantity which appears below, is $\vec{J}=%
\dfrac{1+\gamma +\alpha _{1}/4}{2}\vec{J}_{\oplus }$. We define $J=\left\| 
\vec{J}\right\| \simeq \left\| \vec{J}_{\oplus }\right\| =J_{\oplus }\simeq
145 \, {\rm cm}^{2}$.\newline

$\bullet \qquad \vec{g}=-2\;\dfrac{\vec{J}\wedge \vec{r}}{r^{3}}+\dfrac{1}{2}%
\alpha _{1}U\;\vec{w}$ is the definition of $\vec{g}.$\newline

$\bullet \qquad U$ is the Newtonian potential 
\begin{equation}
U=\dfrac{M_{\oplus }}{r}\left( 1-J_{2}\left( \dfrac{R_{\oplus }}{r}\right)
^{2}P_{2}+\Delta \right) +U_{\ast }  \label{4pole}
\end{equation}%
where $R_{\oplus }$ is the radius of the Earth and $U_{\ast }$ the potential
due to the Moon, the Sun and the planets\footnote{%
An arbitrary constant can always be added to $U_{\ast }.$ It is chosen in
such a way that zero is the mean value of $U_{\ast }$ at point $O$ in the
satellite.}. In spherical coordinates the Legendre polynomial $P_{2}$ reads $%
P_{2}=\dfrac{1}{2}\left( 3\cos ^{2}\theta -1\right) .$ The quadrupole
coefficient $J_{2}$ is of order of $10^{-3}$ and $\Delta $ represents the
higher harmonics; it is of order of $10^{-6}$ \cite{march}. It depends on
the angle $\varphi $ and on the time $t$ because of the rotation of the
Earth.

In the non rotating geocentric coordinates the significant fundamental
element is 
\begin{equation}
ds^{2}=\left( 1-2U\right) dt^{2}+2g_{0k}\,dx^{k}dt-\left( 1+2\gamma U\right)
\delta _{jk}\,dx^{j}dx^{k}  \label{ds2}
\end{equation}
where $\left( \vec{g}\right) _{k}=-\left( \vec{g}\right) ^{k}=g_{0k}.$ In
the expression (\ref{ds2}), we have dropped post Newtonian corrections which
are too small to be considered here.

\subsection{Orders of magnitude}

We consider a nearly free falling, Earth satellite on a nearly circular
orbit of radius $r$ (with $r\simeq 7000 \, {\rm km}).$

\subsubsection{Orbital data.}

The velocity of the satellite is of order $O_{1}\sim \sqrt{\dfrac{M_{\oplus }%
}{r}}$ ($i.e.$ $O_{1}\simeq 2.5\,10^{-5}$ for $r\simeq 7000 \, {\rm km}).$

In the neighborhood of the satellite the potential $U$ is of the order of
the potential of the Earth $U\sim \dfrac{M_{\oplus }}{r}\sim O_{2};$ the
coefficients $g_{0k}$ fulfills the relation $\left| g_{0k}\right| \gtrsim 
\dfrac{J_{\oplus }}{M_{\oplus }^{\;2}}\left( \dfrac{M_{\oplus }}{r}\right)
^{2}\sim 750\,O_{4}>>O_{4}$(\footnote{
It does not mean that each $g_{0k}$ fulfills the relation but that the terms 
$g_{0k}$ are not all negligible compared to $750\,O_{4}.$ }).

The post Newtonian terms that we have neglected in the metric (\ref{ds2})
are of order $O_{4}.$

\subsubsection{The size of the set-up.}

Now we define $d\sim r\,O_{1}$ ($i.e.$ $d\simeq 175 \, {\rm m}$ for $r\simeq
7000 \, {\rm km}).$ \ We assume that the size of the experimental set-up in the
satellite is $X\sim \varepsilon \,d$ where $\varepsilon <1.$ In the sequel
we will consider an atomic Sagnac unit the size of which does not exceed $0.9
 \, {\rm m}$ (then $\varepsilon \sim 5\,10^{-3}).$

\subsubsection{The local acceleration.}

The acceleration, $\left\| \overrightarrow{a}\right\| ,$ can be measured by
accelerometers comoving with the satellite. It is called ''the acceleration
of the satellite relatively to a local inertial frame''. The satellite is
nearly free falling, therefore $\left\| \overrightarrow{a}\right\|
<<a\lesssim O_{2}\times \dfrac{1}{r}.$

Our assumptions are summarized below. The numerical values are obtained for
the Earth with $r\sim 7000 \, {\rm km},$ $M_{\oplus }/r\sim O_{2}\sim
6\,10^{-10}$ and $v/c\sim O_{1}\sim 2.5\,10^{-5}$ ($i.e.$ \ $v\simeq 7.5
 \, {\rm km} \cdot {\rm s}^{-1})$ where $v$ is the velocity of the satellite
relatively to the geocentric frame:

\begin{center}
$%
\begin{tabular}{c}
\hline
\multicolumn{1}{|c|}{\ } \\ 
\multicolumn{1}{|c|}{%
\begin{tabular}{|l|l|l|}
\hline
$M_{\oplus }/r\sim O_{2}\sim v^{2}$ & $X\lesssim \varepsilon
O_{1}r=\varepsilon \,d,$with\ $\ \varepsilon <1$ & $\left\| \overrightarrow{a
}\right\| \ll a\sim O_{2}\times \dfrac{1}{r}$ \\ \hline
\end{tabular}%
} \\ 
\multicolumn{1}{|c|}{\ } \\ 
\multicolumn{1}{|c|}{%
\begin{tabular}{|l|l|l|l|}
\hline
$M_{\oplus }\simeq 4.4 \, {\rm mm}$ & $J\simeq J_{\oplus }\simeq 145 \, {\rm cm}
^{2}$ & $r\simeq 7\,000 \, {\rm km}$ & $\dfrac{J}{M_{\oplus }^{2}}\simeq 750>>1$
\\ \hline
\end{tabular}%
} \\ 
\multicolumn{1}{|c|}{\ } \\ 
\multicolumn{1}{|c|}{%
\begin{tabular}{|c|c|c|c|}
\hline
$O_{1}\sim d/r$ & $O_{2}\sim r\,a$ & $\left| g_{0k}\right| \sim 750\,O_{4}$
& $O_{4}$ \\ \hline
$2.5\,10^{-5}$ & $6\,10^{-10}$ & $3\,10^{-16}$ & $4\,10^{-19}$ \\ \hline
\end{tabular}%
} \\ 
\multicolumn{1}{|c|}{\ } \\ 
\multicolumn{1}{|c|}{%
\begin{tabular}{|l|l|}
\hline
$d\sim 175 \, {\rm m}$ & $a\sim 8 \, {\rm m} \cdot {\rm s}^{-2}$ \\ \hline
\end{tabular}%
} \\ 
\multicolumn{1}{|c|}{\ } \\ \hline
\\ 
Table 1%
\end{tabular}%
$
\end{center}

In the sequel we will assume the preceding relations.

\subsection{The NiZiLi comoving metric}

Before closing this section dedicated to generalities, we give the
expression of the fundamental element associated to the NiZiLi metric.

First we choose an origin, $O,$ in the satellite and we choose a tetrad, $e_{
\hat{\sigma}}^{\alpha }$ whose vector $e_{\hat{0}}^{\alpha }=u^{\alpha }$ is
the 4-velocity, $u^{\alpha },$ of point $O.$ Thus, the vectors $e_{\hat{k}
}^{\alpha }$ define the basis of the space vectors for the observer $O.$ The
coordinates which are associated to the tetrad are the space coordinates $X^{
\hat{k}}$ and the time $X^{\hat{0}}=T$.

Following the procedure defined in \cite{LiZ} and \cite{LiNi} we find the
fundamental element and the local metric tensor

\begin{eqnarray}
ds^{2} &=&G_{\hat{\alpha}\hat{\beta}}\,dX^{\hat{\alpha}}dX^{\hat{\beta}
}+\varepsilon ^{2}\,O_{6}{ \ \ \rm with}  \label{local} \\
&&\;  \nonumber \\
G_{\hat{0}\hat{0}} &=&1+2\,\vec{a}\cdot \vec{X}+\left( \vec{a}\cdot \vec{X}
\right) ^{2}-\left( \vec{\Omega}\wedge \vec{X}\right) ^{2}  \nonumber\\
&&-R_{\hat{0}\hat{k}
\hat{0}\hat{\jmath}}\,X^{\hat{k}}X^{\hat{\jmath}}-\dfrac{1}{3}R_{\hat{0}\hat{
k}\hat{0}\hat{\jmath},\hat{\ell}}X^{\hat{k}}X^{\hat{\jmath}}X^{\hat{\ell}
}+...  \nonumber \\
G_{\hat{0}\hat{m}} &=&\Omega _{\hat{m}\hat{k}}\,X^{\hat{k}}-\dfrac{2}{3}R_{
\hat{0}\hat{k}\hat{m}\hat{\jmath}}\,X^{\hat{k}}X^{\hat{\jmath}}-\dfrac{1}{4}
R_{\hat{0}\hat{k}\hat{m}\hat{\jmath},\hat{\ell}}X^{\hat{k}}X^{\hat{\jmath}%
}X^{\hat{\ell}}+...  \nonumber \\
G_{\hat{n}\hat{m}} &=&\eta _{\hat{n}\hat{m}}-\dfrac{1}{3}R_{\hat{n}\hat{k}%
\hat{m}\hat{\jmath}}X^{\hat{k}}X^{\hat{\jmath}}-\dfrac{1}{6}R_{\hat{n}\hat{k}%
\hat{m}\hat{\jmath},\hat{\ell}}X^{\hat{k}}X^{\hat{\jmath}}X^{\hat{\ell}}+...
\nonumber
\end{eqnarray}%
where we have used vector notations \textit{i.e.} $\overrightarrow{V}$ for $%
\left\{ V^{\hat{k}}\right\} ,\ \overrightarrow{V}\cdot \overrightarrow{W}$
for $\sum V^{\hat{k}}\,W^{\hat{k}},$ etc. Every quantity, except the space
coordinates $X^{\hat{k}},$ are calculated at point $O.$ Thus they are
functions of the time $T.$

$R_{\hat{\alpha}\hat{\beta}\hat{\sigma}\hat{\mu}}$ is the Riemann tensor
obtained from $R_{\alpha \beta \sigma \mu }$ at point $O:$
\begin{equation}
R_{\alpha \beta \sigma \mu }=\Gamma _{\alpha -\beta \mu ,\sigma }-\Gamma
_{\alpha -\beta \sigma ,\mu }+\Gamma _{\beta \sigma }^{\varepsilon }\Gamma
_{\varepsilon -\alpha \mu }-\Gamma _{\beta \mu }^{\varepsilon }\Gamma
_{\varepsilon -\alpha \sigma }  \label{riem}
\end{equation}
where $ \Gamma_{\alpha -\beta \mu}$ is the Christoffel symbol.

$\Omega _{\hat{\jmath}\hat{k}}$ is the antisymmetric quantity
\begin{equation}
\Omega _{\hat{\jmath}\hat{k}}=
\dfrac{1}{2}
\left(
  g_{\hat{0}\hat{\jmath},\hat{k}}-g_{\hat{0}\hat{k},\hat{\jmath}}\right) _{O}
+ \dfrac{1}{2}
\left( 
  \left( e_{\hat{\jmath}}^{\beta }\,\dfrac{de_{\hat{k}}^{\alpha
      }}{ds}-\dfrac{de_{\hat{\jmath}}^{\beta
      }}{ds}\,e_{\hat{k}}^{\alpha }
  \right)
  g_{\alpha \beta }
\right) _{O}  
\label{omega}
\end{equation}

Due to the antisymmetry of $\Omega _{\hat{m}\hat{k}},$ the quantity $\Omega
_{\hat{m}\hat{k}}\,X^{\hat{k}}dX^{\hat{m}}$ which is present in the
expression of $ds^{2}$ can be written as $\Omega _{\hat{m}\hat{k}}\,X^{\hat{k%
}}dX^{\hat{m}}=\left( \overrightarrow{\Omega }_{0}\wedge \overrightarrow{X}%
\right) \cdot d\overrightarrow{X}$. The space vector $\overrightarrow{\Omega 
}_{0}$ is the physical angular velocity. It is measured by gyroscopes tied
to the three space orthonormal vectors $e_{\hat{k}}^{\alpha }$ :

The vector $\overrightarrow{a}$ is the physical acceleration which can be
measured by an accelerometer comoving with $O.$ It is the spatial projection
at point $O$ of the 4-acceleration of point $O.$

At point $O$ ($i.e.$ $\overrightarrow{X}=\overrightarrow{0})$ the time $T$
is the proper time delivered by an ideal clock comoving with $O.$

The first tetrad that we consider is actually called $e_{\hat{\sigma}%
}^{\alpha }:$

\begin{eqnarray}
e_{\hat{0}}^{0} &=&u^{0}=1+\dfrac{\vec{v}^{2}}{2}+U\;+O_{4},\;\;e_{\hat{0}%
}^{k}=u^{k}=\left( 1+\dfrac{\vec{v}^{2}}{2}+U\right) v^{k}+O_{4}  \nonumber \\
e_{\hat{k}}^{0} &=&\left( 1+\dfrac{\vec{v}^{2}}{2}+U\right) \,v^{k}+\gamma
U\,v^{k}-g_{0k}+O_{4}  \label{tetrad} \\
e_{\hat{k}}^{j} &=&\delta _{k}^{j}+\dfrac{1}{2}v^{j}v^{k}+\dfrac{1}{2}\gamma
U\,\delta _{k}^{j}+O_{4}  \nonumber
\end{eqnarray}

Calculating $\vec{\Omega}_{0}$ one finds

\begin{eqnarray}
\vec{\Omega}_{0} &=&\vec{\Omega}_{LT}+\vec{\Omega}_{dS}+\vec{\Omega}_{Th}
\label{omzero} \\
\left( \vec{\Omega}_{LT}\right) ^{\hat{k}} &\simeq &\left( \dfrac{\vec{J}}{%
r^{3}}-\dfrac{3}{r^{3}}\left( \vec{J}\cdot \vec{n}\right) \,\vec{n}\;-\dfrac{%
\alpha _{1}}{4}\vec{\nabla}U\wedge \vec{w}\right) ^{k}  \label{omlt} \\
\left( \vec{\Omega}_{dS}\right) ^{\hat{k}} &\simeq &\left( \left( 1+\gamma
\right) \vec{\nabla}U\wedge \vec{v}\right) ^{k}\ {\rm and}\ \ \ \ \left( 
\vec{\Omega}_{Th}\right) ^{\hat{k}}\simeq \left( \dfrac{1}{2}\vec{v}\wedge 
\dfrac{d\vec{v}}{dt}\right) ^{k}  \label{omdsth}
\end{eqnarray}
$\vec{\Omega}_{LT}$ is the Lense-Thirring angular velocity, $\vec{\Omega}%
_{dS}$\ and\ $\vec{\Omega}_{Th}$ are the de Sitter and the Thomas
terms
\footnote{
The Thomas term reads 
$\vec{\Omega}_{Th}=\dfrac{1}{2}\vec{v}\wedge\vec{A}$ 
where $\vec{A}$ is the ''acceleration''. From
the relativistic point of view, it would be better to define the Thomas term
with the local physical acceleration, $\vec{A}\simeq \dfrac{d\vec{%
v}}{dt}-\vec{\nabla}U$, rather than the acceleration, $\dfrac{d\vec{v}}{dt}$,
relatively to the geocentric frame.
}.

Now it is straightforward to calculate the order of magnitude of the
Lense-Thirring angular velocity. One finds $\dfrac{J}{M_{\oplus }^{\;2}}O_{4}%
\dfrac{c}{r}\sim 10^{-14} \, {\rm rad} \cdot {\rm s}^{-1}.$

We will see that it is relevant to limit the expansion of the metric at
order $\varepsilon ^{2}\,O_{6};$ therefore we consider only the linear
expression of the Riemann tensor above and we neglect he term $\left( \vec{a}%
\cdot \vec{X}\right) ^{2}$ in the metric \ref{local}.

In the sequel we consider an other tetrad $u_{\left( \sigma \right)
}^{\alpha }.$ Except for the notations, the general results above still hold
with the associated coordinates.

\section{Aberration and deflection of the light}

\label{phaseth}

In the satellite, the experimental set-up is tied to a telescope which
points towards a ''fixed'' star. We assume that the star is far enough for
the parallax to be negligible. However it is necessary to account for the\
gravitational deflection of the light ray and for the aberrations in order
to describe the rotational motion of the telescope during the orbital motion
of the satellite.

\begin{figure}
\centering\includegraphics[width =.6\linewidth]{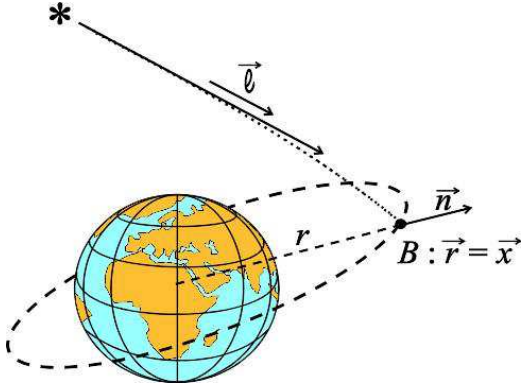}
\caption{\label{figdef} The deflection of light.}
\end{figure}

In space time, the direction of the light from the star is given by the
4-vector $L_{\alpha }=\left\{ 1,\;\dfrac{\partial _{k}\,\varphi }{\partial
_{0}\,\varphi }\right\} $ where $\varphi $ is the phase of the light. In
order to calculate the phase $\varphi (t,x^{k})$ at point $\left\{
x^{k}\right\} $ and time $t$ we use the method which is summarized in annex
A \cite{LetT}.

The monopolar term of the Newtonian potential gives 
\begin{equation}
L_{\alpha }=\left\{ 1,\;-\ell ^{k}+\left( 1+\gamma \right) \dfrac{M_{\oplus }%
}{r}\dfrac{n^{k}-\ell ^{k}}{1-\vec{n}\cdot \vec{\ell}}\right\}  \label{lalfa}
\end{equation}
where $\overrightarrow{\ell }$ is the unitary vector of figure \ref{figdef}
and $\overrightarrow{n}=\overrightarrow{r}/r$ .

Several contributions must be added to the expression (\ref{lalfa}). 
\begin{equation*}
L_{\alpha }=\left\{ 1,\;-\ell ^{k}+\left( 1+\gamma \right) \dfrac{M}{r}%
\dfrac{n^{k}-\ell ^{k}}{1-\vec{n}\cdot \vec{\ell}}+\delta \ell ^{k}+\delta
L^{k}\right\}
\end{equation*}

$\bullet \qquad $The term $\delta \ell ^{k}$ is due to the quadrupolar term
of the Earth. We give its value in appendix A. Its order of magnitude is $%
J_{2}O_{2}\sim 10^{-3}O_{2}$ when $\overrightarrow{\ell }$ is nearly
orthogonal to the plane of the orbit.

$\bullet \qquad $The term due to $\dfrac{1}{2}\alpha _{1}U\,w^{k},$ a part
of $g_{0k}$ in the metric (\ref{ds2}), results in the modification $%
M_{\oplus }\rightarrow M=M_{\oplus }\left( 1-\dfrac{\alpha _{1}\,\vec{w}%
\cdot \vec{\ell}}{2\left( 1+\gamma \right) }\right) .$ The demonstration is
straight forward.

$\bullet \qquad $The contribution due to the rotation of the Earth has
already been considered in the literature (\cite{Ibanez}). Since it is proportional to the ratio of the angular momentum and the square of $r$, it is negligible. The same
conclusion holds for the term $\Delta $ in expression (\ref{4pole}).

$\bullet \qquad $The Sun, the Moon and the other planets, give a
contribution due to $U_{\ast }$ in (\ref{4pole}); it varies slowly with the
time and it is negligible, especially within the framework of a Fourier
analysis at a much higher frequency.

\ 

For the observer $O,$ the space direction of the light is the four vector $%
\lambda ^{\alpha }=L^{\alpha }-L_{\beta }\,u^{\beta }\,u^{\alpha }.$ The
components of $\lambda ^{\alpha }$ relatively to the tetrad $e_{\hat{\alpha}%
}^{\beta }$ are $\lambda _{\hat{\alpha}}=e_{\hat{\alpha}}^{\beta }\lambda
_{\beta }.$ 
\begin{equation*}
\lambda ^{\hat{k}}=-\lambda _{\hat{k}}=-L_{\alpha }e_{\hat{k}}^{\alpha
}\;,\;\;\lambda ^{\hat{0}}=-\lambda _{\hat{0}}=0
\end{equation*}%
In order to calculate $\lambda ^{\hat{k}},$ we use the expression (\ref%
{tetrad}) for the tetrad. We normalize $\vec{\lambda},$ \textit{i.e.} we
define $\vec{\Lambda}=\Lambda \,\vec{\lambda}$ such as $-\Lambda _{\alpha
}\Lambda ^{\alpha }=\vec{\Lambda}\cdot \vec{\Lambda}=1.$ The tetrad (\ref%
{tetrad}) is especially useful to catch the orders of magnitude of the
various terms involved. However it is not the comoving tetrad that we are
looking for because the telescope that points towards the far away star
rotates relatively to this tetrad. The angular velocity of the telescope
relatively to $\left\{ e_{\hat{k}}^{\alpha }\right\} $ is $\vec{\Omega}%
_{\ast }=\vec{\Lambda}\wedge \dfrac{d\vec{\Lambda}}{dt}.$ Straightforward
calculations give%
\begin{eqnarray*}
\Lambda ^{\hat{k}} &=&\ell ^{k}\,\left( 1+\vec{\ell}\cdot \vec{v}-\dfrac{1}{2%
}\vec{v}^{\,2}+\left( \vec{\ell}\cdot \vec{v}\right) ^{2} \right.\\
&&\left.+\left( 1+\gamma
\right) \dfrac{M}{r}\dfrac{\vec{n}\cdot \vec{\ell}}{1-\vec{n}\cdot \vec{\ell}%
}-\vec{\ell}\cdot \left( \delta \vec{\ell}+\delta \vec{L}\right) \right) \\
&&-v^{k}\,\left( 1+\dfrac{1}{2}\left( \vec{\ell}\cdot \vec{v}\right) \right)
-n^{k}\dfrac{M}{r}\dfrac{1+\gamma }{1-\vec{n}\cdot \vec{\ell}}+\delta \vec{%
\ell}+\delta \vec{L}+O_{3}
\end{eqnarray*}%
and%
\begin{eqnarray}
\left( \vec{\Omega}_{\ast }\right) ^{\hat{k}} &=&-\left( \vec{\ell}\wedge 
\dfrac{d\vec{v}}{dt}\right) ^{k}+\left( \vec{v}\wedge \dfrac{d\vec{v}}{dt}%
\right) ^{k} \label{abrot}\\
&&-\dfrac{3}{2}\left( \vec{\ell}\cdot \vec{v}\right) \left( \vec{%
\ell}\wedge \dfrac{d\vec{v}}{dt}\right) ^{k}+\dfrac{1}{2}\left( \vec{\ell}%
\cdot \dfrac{d\vec{v}}{dt}\right) \left( \vec{\ell}\wedge \vec{v}\right) ^{k}
\nonumber \\
&&-\dfrac{M}{r^{2}}\dfrac{1+\gamma }{1-\vec{n}\cdot \vec{\ell}}\left( \left( 
\vec{\ell}\wedge \vec{v}\right) ^{k}+\left( \vec{\ell}\wedge \vec{n}\right)
^{k}\;\left[ \dfrac{\vec{\ell}\cdot \vec{v}\;-\vec{n}\cdot \vec{v}}{1-\vec{n}%
\cdot \vec{\ell}}-\vec{n}\cdot \vec{v}\right] \right)  \nonumber \\
&&+\left( \vec{\ell}\wedge \dfrac{d\delta \vec{\ell}}{dt}\right) ^{k}+\dfrac{%
1}{r}\times O_{4}  \nonumber
\end{eqnarray}%
Let us notice that we neglect the terms of order $\dfrac{1}{r}\times O_{4}$,
which are much smaller than the Lense-Thirring angular velocity ($\sim
750\,O_{4}\times \dfrac{1}{r}).$

Now we introduce the tetrad $u_{\left( \sigma \right) }^{\alpha }$ which is
obtained from $e_{\hat{\rho}}^{\alpha }$ through a pure space rotation (%
\textit{i.e}. $u_{\left( 0\right) }^{\alpha }=e_{\hat{0}}^{\alpha
}=u^{\alpha })$ and whose vector $u_{\left( 1\right) }^{\alpha }$ points
towards the far away star ($u_{\left( 1\right) }^{\alpha }=-\Lambda ^{\alpha
})$. The rotation of the tetrad $\left\{ u_{\left( \sigma \right) }^{\alpha
}\right\} $ relatively to $\left\{ e_{\hat{\rho}}^{\alpha }\right\} $ is
characterized by the most general angular velocity $\vec{\Omega}_{u/e}=\vec{%
\Omega}_{\ast }-\varpi \vec{\Lambda}$ where $-\varpi \vec{\Lambda}$ is an
arbitrary angular velocity around the apparent direction of the star.

\section{The local metric in the satellite}

\label{localmetric}

\subsection{The relevant terms}

Now we need to calculate the NiZiLi metric associated to the tetrad $%
u_{\left( \sigma \right) }^{\alpha }$. We limit the accuracy of our
development to a few tens per cent in order to take advantage of important
simplifications.

Let us assume that the fundamental element is known : 
\begin{equation*}
ds^{2}=\left( 1+K_{\left( 0\right) \left( 0\right) }\right)
dT^{2}+2K_{\left( 0\right) \left( k\right) }dTdX^{\left( k\right) }+\left(
\eta _{\left( k\right) \left( j\right) }+K_{\left( k\right) \left( j\right)
}\right) dX^{\left( k\right) }dX^{\left( j\right) }
\end{equation*}%

We want to calculate the effect of $ K_{\left( \alpha \right) \left( \beta \right) }$
  on the atom interferometer in the satellite.
In order to use the method summarized in the appendix A we calculate the
quantity $\Psi $ which gives the physical quantities that one can measure~:%
\begin{equation}
\Psi =K_{\left( 0\right) \left( 0\right) }+2K_{\left( 0\right) \left(
k\right) }v_{g}^{\left( k\right) }+K_{\left( k\right) \left( j\right)
}v_{g}^{\left( k\right) }v_{g}^{\left( j\right) }  \label{psipsi}
\end{equation}%
where $v_{g}^{\left( k\right) }$ is the velocity of the atoms (\textit{i.e.}
the unperturbed group velocity).

We are dealing with slow cold atoms, therefore $v_{g}\sim \eta O_{1}$ with $%
\eta <<1$ (we can take the value $\eta \sim O_{1}$ valid for current
experiments). In $\Psi ,$ the Lense-Thirring term arises from $2K_{\left(
0\right) \left( k\right) }v_{g}^{\left( k\right) };$ it is of order of $2%
\dfrac{J}{r^{3}}Xv_{g}\sim 1500\,\eta \varepsilon O_{6}.$ If the expansion
of $K_{\left( 0\right) \left( 0\right) }$ is limited to order $\varepsilon
^{2}O_{6},$ the accuracy is of order $\dfrac{\varepsilon }{1500\eta }\sim
13\%$ with $v_{g}\simeq 20 \, {\rm cm} \cdot {\rm s}^{-1}.$

In the case that we consider, the following orders of magnitude hold true : $%
\varepsilon \sim \sqrt{O_{1}}=O_{1/2}$ and $\eta \sim O_{1}.$ Therefore, it
is not necessary to consider terms smaller than $O_{7}\sim \varepsilon
^{2}\,O_{6}$ in $K_{\left( 0\right) \left( 0\right) },$ smaller than $O_{5}$
in $K_{\left( 0\right) \left( k\right) }$ and smaller than $O_{3}$ in $%
K_{\left( k\right) \left( j\right) }.$ The same relations hold in the
coordinate system associated to the tetrad $\left\{ u_{\left( \sigma \right)
}^{\alpha }\right\} $ as well as $\left\{ e_{\hat{\rho}}^{\alpha }\right\} $
because both are next to each other.

It can be proved that in the expression \ref{4pole} the time can be
considered as a parameter and that, within the present approximations, it is
not necessary to consider time derivative in the expression \ref{local}.
Therefore one obtains%
\begin{eqnarray*}
G_{\hat{0}\hat{0}} &=&1+2\,\vec{a}\cdot \vec{X}-\hat{U},_{\hat{k}\hat{\jmath}%
}\,X^{\hat{k}}X^{\hat{\jmath}}-\dfrac{1}{3}\hat{U},_{\hat{k}\hat{\jmath}\hat{%
\ell}}\,X^{\hat{k}}X^{\hat{\jmath}}X^{\hat{\ell}}+\varepsilon ^{2}\,O_{6} \\
G_{\hat{0}\hat{m}} &=&-\left\{ \vec{\Omega}_{0}\wedge \vec{X}\right\} ^{\hat{%
m}}+\varepsilon ^{2}\,O_{5}{\rm  and \ }G_{\hat{n}\hat{m}}=\eta _{\hat{n}%
\hat{m}}+\varepsilon ^{2}O_{4}
\end{eqnarray*}%
where $\vec{\Omega}_{0}$ is given above (see expression \ref{omzero}) while
the expressions such as $\hat{U},_{\hat{k}\hat{\jmath}}$ are nothing but $%
\left( U_{,mn}\,e_{\hat{k}}^{m}e_{\hat{\jmath}}^{n}\right) _{O}.$ The
position of the observer changes with time, therefore this quantity is a
function of $T.$

\ 

In $G_{\hat{0}\hat{0}},$ such an expansion is limited to the terms of order
of $\varepsilon ^{2}\,O_{6},$ thus the approximation $e_{\hat{k}}^{m}=\delta
_{\hat{k}}^{m}$ is valid and $\hat{U},_{\hat{k}\hat{\jmath}}\simeq \left(
U_{,kj}\right) _{O}.$ For $\vec{\Omega}_{0}$ the same holds true~: $\left( 
\vec{\Omega}_{0}\right) ^{\hat{k}}\simeq \left( \vec{\Omega}_{0}\right) ^{k}$
(see expression \ref{omzero}). Therefore, one can identify the space vectors 
$\overrightarrow{e}_{\hat{k}}$ of the tetrad and the space vectors $%
\overrightarrow{\partial }_{k}$ of the natural basis associated to the
geocentric coordinates. This would not be valid with an higher accuracy
where terms smaller than $\varepsilon ^{2}\,O_{6}$ are considered.

The change of the tetrad $u_{\left( \sigma \right) }^{\alpha
}\longleftrightarrow e_{\hat{\rho}}^{\alpha }$ is just an ordinary change of
basis in the space of the observer $O.$ In this transformation, $dT,$ $G_{%
\hat{0}\hat{0}}=G_{\left( 0\right) \left( 0\right) }$, $G_{\hat{0}\hat{m}%
}dX^{\hat{m}}=G_{\left( 0\right) \left( k\right) }dX^{\left( k\right) }$ and 
$G_{\hat{n}\hat{m}}dX^{\hat{m}}dX^{\hat{n}}=G_{\left( j\right) \left(
k\right) }dX^{\left( j\right) }X^{\left( k\right) }$ behave as scalars.

As a consequence, we put forward that within the required accuracy, it is
possible to give a\ simple description of the Hyper project with the
Newtonian concept of space. It is now straightforward to calculate $\Psi $ :%
\begin{eqnarray*}
\Psi &=&2\,\vec{a}\cdot \vec{X}-\hat{U},_{\left( k\right) \left( j\right)
}\,X^{\left( k\right) }X^{\left( j\right) }-\dfrac{1}{3}\hat{U},_{\left(
k\right) \left( j\right) \left( \ell \right) }\,X^{\left( k\right)
}X^{\left( j\right) }X^{\left( \ell \right) } \\
&&-2\underset{\left( k\right) }{\sum }\left\{ \left( \vec{\Omega}_{0}+\vec{%
\Omega}_{\ast }\right) \wedge \vec{X}\right\} ^{\left( k\right)
}v_{g}^{\;\left( k\right) }+\varepsilon ^{2}O_{6}
\end{eqnarray*}

Now we assume that any quantity can be known with an accuracy 
%TCIMACRO{\U{b5} }%
%BeginExpansion
$\mu$
%EndExpansion
of order of $10^{-5}\lesssim O_{1}.$ It means that the geometry of the
experimental device is known with an accuracy of order of $10 {\rm %
%TCIMACRO{\U{3bc}}%
%BeginExpansion
\mu%
%EndExpansion
m}.$ The position of the point $O$ on the trajectory of the satellite is
known with an accuracy better than $70 \, {\rm m},$ the eccentricity can be
controlled to be smaller than $O_{1},$ etc.

We consider that $\Psi $ is the amount of two terms, $\Psi _{k}$ and $\Psi
_{u}~:$ the term $\Psi _{k}$ can be modelled with the required accuracy
while $\Psi _{u}$ is unknown. The terms $\Psi _{k}$ fulfills the condition 
%TCIMACRO{\U{b5}}%
%BeginExpansion
$\mu$%
%EndExpansion
$\times \Psi _{k}\lesssim \varepsilon ^{2}\,O_{6}.$ With the previous orders
of magnitude ($\varepsilon \sim O_{1/2}\sim 5\,10^{-3}$ and 
%TCIMACRO{\U{b5}}%
%BeginExpansion
$\mu$%
%EndExpansion
$\,\sim O_{1}\sim 2.5\,10^{-5})$ one finds%
\begin{equation*}
\Psi _{u}=2\,\vec{a}\cdot \vec{X}-\hat{U},_{\left( k\right) \left( j\right)
}\,X^{\left( k\right) }X^{\left( j\right) }-2\left\{ \vec{\Omega}\wedge \vec{%
X}\right\} \cdot \overrightarrow{v}_{g}+\varepsilon ^{2}O_{6}/%
%TCIMACRO{\U{b5}}%
%BeginExpansion
{\mu}%
%EndExpansion
\end{equation*}%
$\hat{U},_{\left( k\right) \left( j\right) }$ is obtained from the
expression \ref{4pole} of $U$ limited to the monopolar and the quadrupolar
terms and 
\begin{equation*}
\left( \vec{\Omega}\right) ^{\left( k\right) }=\left( \vec{\Omega}%
_{LT}-\varpi \overrightarrow{\Lambda }-\overrightarrow{\Lambda }\wedge 
\dfrac{d\overrightarrow{v}}{dt}\right) ^{\left( k\right) }+\dfrac{O_{3}}{r}
\end{equation*}%
where $\vec{\Omega}_{LT}$ is obtained from \ref{omlt} and $\dfrac{d%
\overrightarrow{v}}{dt}\simeq \overrightarrow{a}-\overrightarrow{\nabla }%
U\simeq \overrightarrow{a}+\dfrac{M_{\oplus }}{r^{2}}\overrightarrow{n}.$

\subsection{Calculation of $\Psi _{u}$}

We consider that the motion of the satellite is a Newtonian motion which
takes place in the $\left( x,y\right) $-plane while the vector $%
\overrightarrow{\ell }$ lies in the $\left( x,z\right) $-plane. We assume
that the eccentricity, $e,$ does not exceed $O_{1}$.

We define

\begin{eqnarray*}
\vec{J} &=&J_{x}\,\vec{e}_{\hat{1}}+J_{y}\,\vec{e}_{\hat{2}}+J_{z}\,\vec{e}_{%
\hat{3}}\;,\;\;\vec{n}=\cos \theta \,\vec{e}_{\hat{1}}+\sin \theta \,\vec{e}%
_{\hat{2}} \\
-\vec{\ell} &=&\cos \alpha \,\vec{e}_{\hat{1}}+\sin \alpha \,\vec{e}_{\hat{3}%
}\;,\;\;\vec{w}=w_{x}\,\vec{e}_{\hat{1}}+w_{y}\,\vec{e}_{\hat{2}}+w_{z}\,%
\vec{e}_{\hat{3}}
\end{eqnarray*}
$\vec{J},$ $\vec{\ell}$ and $\vec{w}$ are constant vectors. The angle $%
\theta $ and the distance $r$ depend on the time $T$.

\begin{figure}
\centering\includegraphics[width =.6\linewidth]{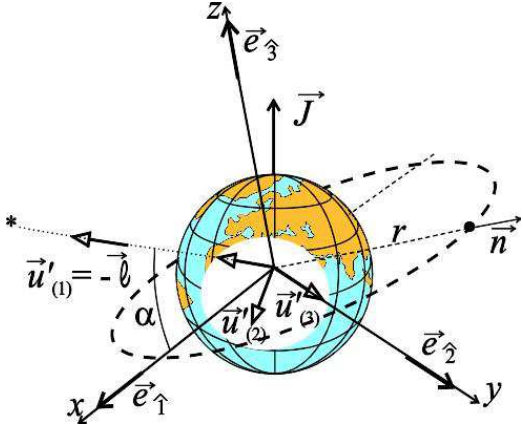}
\caption{\label{short3} The satellite and the fixed star.}
\end{figure}

First we define the special triad $\vec{u}_{(n)}^{\prime }$~: 
\begin{equation*}
\vec{u}_{(1)}^{\prime }=\cos \alpha \,\overrightarrow{e}_{\hat{1}}+\sin
\alpha \overrightarrow{e}_{\hat{3}},\;\;\vec{u}_{(2)}^{\prime }=\sin \alpha
\,\overrightarrow{e}_{\hat{1}}-\cos \alpha \overrightarrow{e}_{\hat{3}},\;\;%
\vec{u}_{(3)}^{\prime }=\overrightarrow{e}_{\hat{2}}
\end{equation*}

Let us outline that we have defined $\left( \vec{u}_{(1)}^{\prime }\right) ^{%
\hat{k}}=-\left( \vec{\Lambda}\right) ^{\hat{k}}+O_{1}$. Then, in order to
obtain the final tetrad $u_{\left( \sigma \right) }^{\alpha },$ we perform
an arbitrary rotation around $\vec{\Lambda}~:$ 
\begin{equation}
\begin{tabular}{l}
$\vec{u}_{(1)}=-\vec{\Lambda}=\vec{u}_{(1)}^{\prime }+O_{1}$ \\ 
$\vec{u}_{(2)}=\vec{u}_{(2)}^{\prime }\cos \sigma +\vec{u}_{(3)}^{\prime
}\sin \sigma +O_{1}$ \\ 
$\vec{u}_{(3)}=-\vec{u}_{(2)}^{\prime }\sin \sigma +\vec{u}_{(3)}^{\prime
}\cos \sigma +O_{1}$%
\end{tabular}
\label{uprimeau}
\end{equation}%
where $-\dfrac{d\sigma }{dT}=-\varpi $ is the angular velocity of the triad $%
\left\{ \vec{u}_{(k)}\right\} $ relatively to $\left\{ \vec{u}_{(k)}^{\prime
}\right\} .$

We can now assume that the experimental set-up is comoving with the triad $%
\vec{u}_{(n)}$ whose vector $\vec{u}_{(1)}$ points towards the fixed star.

In order to calculate the quadrupolar term in $U,$ we define the unitary
vector, $\overrightarrow{k}$ along the axi-symmetry axis. One obtains 
\begin{equation*}
U=\dfrac{M_{\oplus }}{r}-\dfrac{1}{2}J_{2}\dfrac{M_{\oplus }}{r}\left( 
\dfrac{R_{\oplus }}{r}\right) ^{2}\left( 3\,\left( \overrightarrow{k}\cdot 
\overrightarrow{n}\right) ^{2}-1\right)
\end{equation*}

Therefore $\Psi _{u}$ reads

\begin{equation}
\begin{array}{ll|l}
\Psi _{u}= & -2\left( \left( \dfrac{\vec{J}}{r^{3}}-\dfrac{3\left( \vec{J}\cdot \vec{n}\right) }{r^{3}}\vec{n}+\dfrac{\alpha _{1}\;M_{\oplus }}{
4\,r^{2}}\,\vec{n}\wedge \vec{w}\right) \wedge \vec{x}\right) \cdot \vec{v}
_{g} & A \\ 
& -2\varpi \,\left( \vec{u}_{(1)}\wedge \vec{x}\right) \cdot \vec{v}_{g} & 
B \\ 
& -2\dfrac{M_{\oplus }}{r^{2}}\left( \left( \vec{u}_{(1)}\wedge
    \vec{n}\right) \wedge \vec{x}\right) \cdot \vec{v}_{g}+2\left(
\left( \vec{u}_{(1)}\wedge \vec{a}\right) \wedge \vec{x}\right) \cdot \vec{v}
_{g} & C \\ 
& +\dfrac{M_{\oplus }}{r^{3}}\left( \vec{x}^{2}-3\,\left( \vec{n}\cdot \vec{
x}\right) ^{2}\right) +\dfrac{3}{2}J_{2}\dfrac{M_{\oplus }}{r^{3}}\left( 
\dfrac{R_{\oplus }}{r}\right) ^{2}Q & D \\ 
& +2\,\vec{a}\cdot \vec{x} & E
\end{array}
\label{psiu}
\end{equation}

with 
\begin{eqnarray*}
Q &=&\left( 1-5\,\left( \overrightarrow{k}\cdot \overrightarrow{n}\right)
^{2}\right) \overrightarrow{X}^{\,2}-20\left( \overrightarrow{k}\cdot 
\overrightarrow{n}\right) \left( \overrightarrow{n}\cdot \overrightarrow{X}%
\right) \left( \overrightarrow{k}\cdot \overrightarrow{X}\right) 
\\
&&+5\left( 7\,\left( \overrightarrow{k}\cdot \overrightarrow{n}\right)
^{2}-3\right) \left( \overrightarrow{n}\cdot \overrightarrow{X}\right)
^{2}+2\left( \overrightarrow{k}\cdot \overrightarrow{X}\right) ^{2}\\
&\text{and}&\\
\overrightarrow{J} &=&J\,\overrightarrow{k}
\end{eqnarray*}

In $\Psi _{u},$ the two terms $2\,\vec{a}\cdot \vec{X}$ and $-2\varpi
\,\left( \vec{u}_{(1)}\wedge \overrightarrow{X}\right) \cdot \vec{v}_{g}$
are directly calculated in the local coordinates associated to the tetrad $%
u_{\left( \sigma \right) }^{\alpha }$ moreover, to calculate the other
quantities where $\vec{n}$ is involved, one can drop the terms of order $%
O_{1}$ in the expression of $\vec{u}_{(1)}.$ The reason is that the
quantities that we drop are either negligible or included in $\Psi _{k}.$
Therefore, it is clear that we can consider the space as the ordinary space
of Newtonian physics and that the usual formulae to change the basis $%
\overrightarrow{\partial }_{k}$ into $\overrightarrow{e}_{\hat{k}}$ or $%
\overrightarrow{u}_{\left( k\right) }$are valid.

In (\ref{psiu}), the terms of lines $A,$ $B$ and $C$ are due to various
rotations~: respectively the Lense-Thirring rotation, the spin around the
view line of the star and the aberration. The term of line $D$ gives the
gravitational tidal effects (which are mainly due to the Earth) and the last
term (line $E)$ corresponds to some residual acceleration due to the fact
that point $O$ is not exactly in free fall.

It is now possible to calculate explicitly $\Psi _{u}$ with the coordinates $%
X^{(k)},$ comoving with the experimental set-up
\footnote[8]{
Let us emphasize that the de Sitter
 and the Thomas  angular velocities are not involved in $\Psi_{u}$ but in $\Psi_{K}$.}.

\section{The experimental set-up}

\label{asu}

\subsection{The atomic Sagnac unit}

\begin{figure}
\centering\includegraphics[width =.6\linewidth]{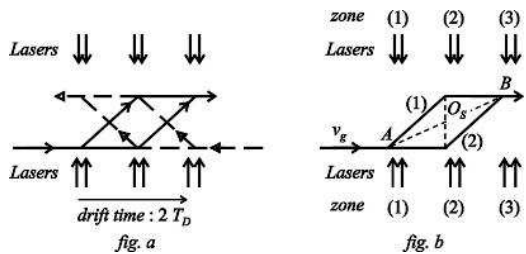}
\caption{\label{figasu} An atomic Sagnac unit (ASU).}
\end{figure}

An atomic Sagnac unit (ASU) is made of two counter-propagating atom
interferometers which discriminate between rotation and acceleration (see
figure \ref{figasu}-$a$).

Each interferometer is a so-called Ramsey-Bord\'{e} interferometer with a
Mach-Zehnder geometry ( figure \ref{figasu}-$b$). The atomic beam from a
magneto-optical trap interacts three times with a laser field. In the first
interaction zone the atomic beam is split coherently, by a Raman effect,
into two beams which are redirected and recombined in the second and the
third interaction zone.

The mass of the atom depends on its internal state, therefore it is not a
constant along the different paths. However, the change of the mass is very
small; it leads to negligible corrections on the main effects which is
already very small. Considering cesium, we assume that the mass of the atom
is a constant $m=133\times 1.66\times 10^{-27}=2.2\times 10^{-25} \, {\rm kg}.$

In this case the wave length of the lasers is $\lambda =850 \, {\rm nm}.$ The
momentum transferred to the atom during the interaction is $\dfrac{4\pi
\hbar }{\lambda }.$ The recoil of the atom results in a Sagnac loop which
permits to measure the angular velocity of the set-up relatively to a local
inertial frame. The device is also sensitive to the accelerations.

One can easily imagine that there are many difficulties to overcome if the
Lense-Thirring effect is to be observed. In particular, the geometrical
constraints appear to be crucial. In an ideal set-up the two interferometers
are identical coplanar parallelograms with their center $O_{S}$ and $O_{S}^{\prime
}$ at the same point. One can consider several perturbation to this
geometrical scheme.

\begin{enumerate}
\item A shift : $O_{S}$ and $O_{S}^{\prime }$ are no longer at the same
point;

\item A tilt : The plane of the two interferometers are now different;

\item A deformations : the interferometers are no longer identical lozenges,
they are no longer parallelogram, even not plane interferometers.
\end{enumerate}

The geometry of the device is fully determine by the interaction between the
initial atomic beam and the lasers~; moreover the geometrical description is
already an idealized model~; Therefore a full treatment of the atom-laser
interaction in a gravitational field is obviously necessary to study the
response of the ASU (see \cite{Antoine}). However the geometrical model is useful to
give a physical intuition of the phenomena. In this context we study here
the effect of a shift on the signal.

In the sequel we consider an ideal set-up where the beams are perfectly
coherent and perfectly parallel, point $O_{S}$ being a perfect center of
symmetry for the atomic paths. We assume also that the two
counter-propagating atom interferometers are identical and located in the
same plane but that the separation of their center of symmetry, $O_{S}$ and $%
O_{S}^{\prime },$ is the vector $\vec{\delta}=\overrightarrow{%
O_{S}O_{S}^{\prime }}.$

We will assume that the velocity, $v_{g},$ of the atoms is of the order of $%
20 \, {\rm cm} \cdot {\rm s}^{-1}$ and that the size of the ASU is of order of $60%
 \, {\rm cm},$ therefore the drift time is $2T_{D}\sim 3 \, {\rm s}.$

\subsection{The phase differences}

\begin{figure}
\centering\includegraphics[width =.6\linewidth]{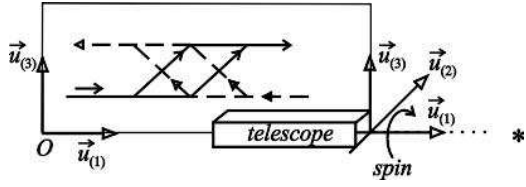}
\caption{\label{figfin} The experimental set-up.}
\end{figure}

The configuration which is presently considered in the Hyper project is on
figure \ref{figfin}.

We consider one among the four interferometers of the device above. Using
the results of appendix A, we obtain the phase difference that we want to
measure: 
\begin{equation}
\delta \varphi =\dfrac{\omega }{2}\int_{t-2T_{D}/(2)}^{t}\psi \,(t^{\prime
})\;dt^{\prime }-\dfrac{\omega }{2}\int_{t-2T_{D}/\left( 1\right) }^{t}\psi
\,(t^{\prime })\;dt^{\prime }  \label{deltafi}
\end{equation}%
The integrals are performed along path (2) and (1) of figure \ref{figasu}-$b$%
. The ''angular frequency'' $\omega $ is defined as $\dfrac{m\,c^{2}}{\hbar }%
.$

We consider that the origin is the center of symmetry $O_{S}.$ The
coordinates are $\overrightarrow{X}=\left\{ X^{\left( k\right) }\right\} .$
We define $\vec{v}_{g}=\dfrac{d\overrightarrow{X}}{dt}$ as the unperturbed
group velocity of the atoms

We make use of the properties of the Newtonian trajectory~: 
\begin{equation*}
r=\dfrac{r_{0}}{1+e\cos \theta },\;\vec{n}=\cos \theta \,\vec{e}_{\hat{1}%
}+\sin \theta \,\vec{e}_{\hat{2}},\;\dfrac{d\theta }{dT}=\Omega _{S}\left(
1+e\cos \theta \right) ^{2}
\end{equation*}%
with $\Omega _{S}=\dfrac{M_{\oplus }}{r_{0}^{\;2}}.$ We consider the case
where the ellipticity, $e,$ is much smaller than unity. Therefore, up to
first order relatively to $e,$ one obtains 
\begin{equation*}
\dfrac{1}{r}=\dfrac{1}{r_{0}}\left( 1+e\cos \theta \right) \;,\;\;\theta
\simeq \Omega _{S}\,t+\theta _{0}+e\sin \left( \Omega _{S}\,t+\theta
_{0}\right)
\end{equation*}%
We assume $e\lesssim O_{1},$ therefore in the expression (\ref{psiu}) of $%
\Psi _{u}$ one can assume that $r=r_{0}$ is a constant because the
corrections due to the eccentricity are included in $\Psi _{k}.$

During the flight time of an atom, one cannot consider that $\overrightarrow{%
n}$ remains a constant in expression (\ref{psiu}). One can consider that the
coordinate, $X=X^{\left( 1\right) }$ of the atom is a function of the time: $%
X=v_{g}\left( T-T_{0}\right) .$ Then $\overrightarrow{n}=\overrightarrow{n}%
_{0}+\delta \overrightarrow{n}$ where $\overrightarrow{n}_{0}$ depends on $%
T_{0}$ only and $\delta \overrightarrow{n}$ on $T_{0}$ and $X/v_{g}.$
Therefore, in expression (\ref{deltafi}), we expand the different terms
relatively to $X/v_{g}.$ Any of the correction can be included in $\Psi _{k}$
except the first order correction, $\delta \psi _{D},$ of the first term in
line $D.$ 
\begin{equation*}
\delta \psi _{D}=-\dfrac{6M_{\oplus }}{r^{4}}\left( \overrightarrow{x}\cdot 
\dfrac{\overrightarrow{v}}{v_{g}}\right) \left( \overrightarrow{x}\cdot 
\overrightarrow{n}_{0}\right) x
\end{equation*}%
The result of the integration is now straightforward. We change the
notation~: $\overrightarrow{n}_{0}\rightarrow \overrightarrow{n}$ and $%
T_{0}\rightarrow T$. In usual units one finds~:%
\begin{equation*}
\begin{tabular}{ll|l}
$\delta \varphi =$ & $-2\dfrac{mc}{\hbar \,r}S\,\left( \dfrac{\vec{J}}{r^{2}}%
-\dfrac{3\left( \vec{J}\cdot \vec{n}\right) }{r^{2}}\vec{n}+\dfrac{\alpha
_{1}\;M_{\oplus }}{4\,r}\,\vec{n}\wedge \vec{w}\right) \cdot \vec{u}_{(2)}$
& $A$ \\ 
& $-2\dfrac{mc}{\hbar r}S\,\dfrac{M_{\oplus }}{r}\,\left( \vec{u}%
_{(1)}\wedge \vec{n}\right) \cdot \vec{u}_{(2)}-2\dfrac{mc}{\hbar }S\left( 
\vec{u}_{(1)}\wedge \dfrac{\overrightarrow{a}}{c^{2}}\right) \cdot \vec{u}%
_{(2)}$ & $C$ \\ 
& $-\dfrac{4\pi \left( cT_{D}\right) ^{2}}{\lambda }\,\left( \vec{u}%
_{(3)}\cdot \dfrac{\vec{a}_{O_{S}}}{c^{2}}\right) $ & $E$ \\ 
&\hspace{-7mm}$-\dfrac{mc}{2\,\hbar r}S\,\dfrac{\left( cT_{D}\right) ^{2}}{r^{2}}\dfrac{%
M_{\oplus }}{r}\;\left( \left( \vec{u}_{(1)}\cdot \vec{n}%
\right) \left( \vec{u}_{(3)}\cdot \dfrac{\overrightarrow{v}}{c}\right)
+\left( \vec{u}_{(3)}\cdot \vec{n}\right) \left( \vec{u}_{(1)}\cdot \dfrac{%
\overrightarrow{v}}{c}\right) \right) $ & $F$%
\end{tabular}%
\end{equation*}%
where $S=\dfrac{4\pi \hbar }{\lambda \,m}v_{g}T_{D}^{\;2}$ is the area of
the Sagnac loop.

Line $B$ in (\ref{psiu}) gives no contribution because $\overrightarrow{X},$ 
$\vec{u}_{(1)}$ and $\overrightarrow{v}_{g}$ are in the same plane. The
quadratic terms have disappeared because $O_{S}$ is a center of symmetry.

The two interferometers of the same ASU are assumed to lie in the same plane
but not necessarily with their center of symmetry $O_{S}$ and $O_{S}^{\prime
}$ at the same point. Therefore adding and subtracting the phase differences
delivered by the two interferometers one finds the two basic quantities
which are measured by the set-up \textit{i.e}.~:

\begin{eqnarray*}
%TCIMACRO{\U{b5}}%
%BeginExpansion
{\mu}%
%EndExpansion
_{1} &=&\dfrac{1}{2}\left( \delta \varphi ^{\prime }-\delta \varphi \right) =%
\dfrac{8\pi }{\lambda }\left( cT_{D}\right) ^{2}\left\{ \dfrac{M_{\oplus }}{%
r^{2}}\,\vec{u}_{(3)}\cdot \vec{n}+\vec{u}_{(3)}\cdot \overrightarrow{a}+%
\vec{\Omega}_{LT}\cdot \vec{u}_{(2)}\right\} \dfrac{v_{g}}{c} \\
&&+\dfrac{2\pi \left( cT_{D}\right) ^{2}}{\lambda }\left\{%
\left( \dfrac{\vec{a}_{O_{S}}}{c^{2}}-\dfrac{\,\vec{a}_{O_{S}^{\prime }}}{%
c^{2}}\right) \cdot \vec{u}_{(3)}\right\} \\
&&-\dfrac{mc}{2\hbar }\dfrac{\left( cT_{D}\right) ^{2}S}{r^{3}}\dfrac{%
M_{\oplus }}{r}\left( \left( \vec{u}_{(1)}\cdot \vec{n}%
\right) \left( \vec{u}_{(3)}\cdot \dfrac{\overrightarrow{v}}{c}\right)
+\left( \vec{u}_{(3)}\cdot \vec{n}\right) \left( \vec{u}_{(1)}\cdot \dfrac{%
\overrightarrow{v}}{c}\right) \right) \\
%TCIMACRO{\U{b5}}%
%BeginExpansion
{\mu}%
%EndExpansion
_{2} &=&\dfrac{1}{2}\left( \delta \varphi ^{\prime }+\delta \varphi \right)
=-\dfrac{4\pi \left( cT_{D}\right) ^{2}}{\lambda }\left\{ \dfrac{%
\overrightarrow{a}}{c^{2}}\cdot \vec{u}_{(3)}\right\}
\end{eqnarray*}%
where now $\overrightarrow{a}=\vec{a}_{O_{S}}+\,\vec{a}_{O_{S}^{\prime }}.$
In the expression above, we have dropped the term $-\dfrac{mc}{2\,\hbar r}S\,%
\dfrac{\left( cT_{D}\right) ^{2}}{r^{2}}\dfrac{M_{\oplus }}{r}\,\dfrac{4 \pi \hbar}{\lambda m v_{g}} \left( 
\vec{u}_{(3)}\cdot \vec{n}\right) \left( \vec{u}_{(3)}\cdot \overrightarrow{v%
}\right) $  because it can be included in $\Psi _{k}.$

\ 

$a_{O_{S}}^{\;\left( k\right) }u_{\left( k\right) }^{\alpha }$ is the
4-acceleration of point $O_{S}.$ Therefore, one calculates $\left( \vec{a}%
_{O_{S}}-\,\vec{a}_{O_{S}^{\prime }}\right) \cdot \vec{u}_{(3)}=-\hat{U}%
_{,\left( k\right) \left( 3\right) }x_{O_{S}}^{\;\left( k\right) }$ where we
have dropped the non significant terms which can be included in $\Psi _{k}.$

The quantities which can be measured are 
\begin{eqnarray*}
\mu_{1}+2v_{g} \mu_{2} 
&=&\dfrac{8\pi }{\lambda }\left( cT_{D}\right) ^{2}\left\{ \dfrac{%
M_{\oplus }}{r^{2}}\vec{u}_{(3)}\cdot \vec{n}+\vec{\Omega}_{LT}\cdot \vec{u}%
_{(2)}\right\} \dfrac{v_{g}}{c} \\
&&\hspace{-25mm}-\dfrac{2\pi \left( cT_{D}\right) ^{2}}{\lambda r}\dfrac{M_{\oplus }}{r}%
\left( \vec{u}_{(3)}\cdot \dfrac{\overrightarrow{\delta }}{r}-3\left( \vec{n}%
\cdot \vec{u}_{(3)}\right) \left( \vec{n}\cdot \dfrac{\overrightarrow{\delta 
}}{r}\right) \right) \\
&&\hspace{-25mm}-\dfrac{2\pi }{\lambda }\left( cT_{D}\right) ^{2}\left( \dfrac{v_{g}T_{D}}{%
r}\right) ^{2}\dfrac{M_{\oplus }}{r^{2}}\left\{ \left( \vec{u}_{(1)}\cdot 
\vec{n}\right) \left( \vec{u}_{(3)}\cdot \dfrac{\overrightarrow{v}}{c}%
\right) +\left( \vec{u}_{(3)}\cdot \vec{n}\right) \left( \vec{u}_{(1)}\cdot 
\dfrac{\overrightarrow{v}}{c}\right) \right\} \\
%TCIMACRO{\U{b5}}%
%BeginExpansion
{\mu}%
%EndExpansion
_{2} &=&\dfrac{1}{2}\left( \delta \varphi ^{\prime }+\delta \varphi \right)
=-\dfrac{4\pi \left( cT_{D}\right) ^{2}}{\lambda }\left\{ \dfrac{%
\overrightarrow{a}}{c^{2}}\cdot \vec{u}_{(3)}\right\}
\end{eqnarray*}%
where $\overrightarrow{\delta }=\overrightarrow{x}_{O_{S}^{\prime }}-%
\overrightarrow{x}_{O_{S}}.$

\subsection{Discussion}

We define the projection, $\overrightarrow{J}_{\shortparallel }$ of $%
\overrightarrow{J}$ on the plane of the orbit~: $\overrightarrow{J}%
_{\shortparallel }=J_{\shortparallel }\left( \cos \theta _{J}\;%
\overrightarrow{e}_{\hat{1}}+\sin \theta _{J}\;\overrightarrow{e}_{\hat{2}%
}\right) $ and $\overrightarrow{w}_{\shortparallel }=w_{\shortparallel
}\left( \cos \theta _{w}\;\overrightarrow{e}_{\hat{1}}+\sin \theta _{w}\;%
\overrightarrow{e}_{\hat{2}}\right) .$ Going back to usual units (with $%
M_{\oplus }\simeq 4.4 \, {\rm mm}$ and $J\simeq 145 \, {\rm cm}^{2})$ we obtain%
\begin{eqnarray*}
%TCIMACRO{\U{b5}}%
%BeginExpansion
{\mu}%
%EndExpansion
_{1}+2\dfrac{v_{g}}{c}%
%TCIMACRO{\U{b5}}%
%BeginExpansion
{\mu}%
%EndExpansion
_{2} &=&\dfrac{2\pi \left( cT_{D}\right) ^{2}}{\lambda r_{0}}\times \left\{
K_{0}+K_{\sigma }+K_{2\sigma }+K_{2\theta }\right.  \\
&&\left. +K_{\theta -\sigma }+K_{\theta +\sigma }+K_{2\theta -\sigma
}+K_{2\theta +\sigma }+K_{2\theta -2\sigma }+K_{2\theta +2\sigma }\right\} 
\end{eqnarray*}%
with%
\begin{eqnarray*}
K_{0} &=&\dfrac{M_{\oplus }}{4\,r_{0}}\left( 3\sin ^{2}\alpha -1\right)
\times \dfrac{\delta ^{(3)}}{r_{0}} \\
K_{\sigma } &=&\dfrac{v_{g}}{c}\left\{ \left[ \left( 1-\sin \alpha \right)
\cos \left( \sigma +\theta _{J}\right) -\left( 1+\sin \alpha \right) \cos
\left( \sigma -\theta _{J}\right) \right] \times \dfrac{J_{\shortparallel }}{%
r_{0}^{\;2}}\right.  \\
&&\left. -4\cos \alpha \,\cos \sigma \times \dfrac{J^{\hat{3}}}{r_{0}^{\;2}}%
\right\} -\dfrac{3M_{\oplus }}{2r_{0}}\cos \alpha \sin \alpha \sin \sigma
\times \dfrac{\delta ^{\left( 1\right) }}{r_{0}} \\
K_{2\sigma } &=&\dfrac{3M_{\oplus }}{4r_{0}}\left( 1-\sin ^{2}\alpha \right)
\,\left[ \sin \left( 2\sigma \right) \times \dfrac{\delta ^{(2)}}{r_{0}}%
+\cos \left( 2\sigma \right) \times \dfrac{\delta ^{(3)}}{r_{0}}\right]  \\
K_{2\theta } &=&-\dfrac{3M_{\oplus }}{4r_{0}}\left( 1-\sin ^{2}\alpha
\right) \,\cos \left( 2\theta \right) \times \dfrac{\delta ^{\left( 3\right)
}}{r_{0}}
\end{eqnarray*}%
\begin{eqnarray*}
K_{\theta -\sigma } &=&\dfrac{M_{\oplus }}{r_{0}}\dfrac{v_{g}}{c}\times
\left\{ -2\left( 1+\sin \alpha \right) \,\sin \left( \theta -\sigma \right) 
\right.  \\
&&\left. +\dfrac{\alpha _{1}}{2}\cos \alpha \,\sin \left( \theta -\sigma
-\theta _{w}\right) \times \dfrac{w_{\shortparallel }}{c}\right.  \\
&&\left. +\dfrac{\alpha _{1}}{2}\left( 1+\sin \alpha \right) \sin \left(
\theta -\sigma \right) \times \dfrac{w^{\hat{3}}}{c}\right\}  \\
K_{\theta +\sigma } &=&\dfrac{M_{\oplus }}{r_{0}}\dfrac{v_{g}}{c}\times
\left\{ -2\left( 1-\sin \alpha \right) \,\sin \left( \theta +\sigma \right) 
\right.  \\
&&\left. +\dfrac{\alpha _{1}}{2}\cos \alpha \,\sin \left( \theta +\sigma
-\theta _{w}\right) \times \dfrac{w_{\shortparallel }}{c}\right.  \\
&&\left. -\dfrac{\alpha _{1}}{2}\left( 1-\sin \alpha \right) \sin \left(
\theta +\sigma \right) \times \dfrac{w^{\hat{3}}}{c}\right\} 
\end{eqnarray*}%
\begin{eqnarray*}
K_{2\theta -\sigma } &=&-3\dfrac{v_{g}}{c}\left( 1+\sin \alpha \right) \cos
\left( 2\theta -\sigma -\theta _{J}\right) \times \dfrac{J_{\shortparallel }%
}{r_{0}^{\;2}} \\
&&+\dfrac{3M_{\oplus }}{4r_{0}}\cos \alpha \,\left( 1+\sin \alpha \right)
\sin \left( 2\theta -\sigma \right) \times \dfrac{\delta ^{\left( 1\right) }%
}{r_{0}} \\
&&-\left( \dfrac{M_{\oplus }}{r_{0}}\right) ^{3/2}\dfrac{cv_{g}T_{D}^{\;2}}{%
2\,r_{0}^{\;2}}\left( 1+\sin \alpha \right) \cos \left( 2\theta -\sigma
\right)  \\
K_{2\theta +\sigma } &=&3\dfrac{v_{g}}{c}\left( 1-\sin \alpha \right) \cos
\left( 2\theta +\sigma -\theta _{J}\right) \times \dfrac{J_{\shortparallel }%
}{r_{0}^{\;2}} \\
&&+\dfrac{3M_{\oplus }}{4r_{0}}\cos \alpha \,\left( 1-\sin \alpha \right)
\sin \left( 2\theta +\sigma \right) \times \dfrac{\delta ^{\left( 1\right) }%
}{r_{0}} \\
&&-\left( \dfrac{M_{\oplus }}{r_{0}}\right) ^{3/2}\dfrac{cv_{g}T_{D}^{\;2}}{%
2\,r_{0}^{\;2}}\left( 1-\sin \alpha \right) \cos \left( 2\theta +\sigma
\right) 
\end{eqnarray*}%
\begin{eqnarray*}
K_{2\theta -2\sigma } &=&\dfrac{3M_{\oplus }}{8\,r_{0}}\left( 1+\sin \alpha
\right) ^{2}\left\{ \sin \left( 2\theta -2\sigma \right) \times \dfrac{%
\delta ^{\left( 2\right) }}{r_{0}}-\cos \left( 2\theta -2\sigma \right)
\times \dfrac{\delta ^{\left( 3\right) }}{r_{0}}\right\}  \\
K_{2\theta +2\sigma } &=&-\dfrac{3M_{\oplus }}{8\,r_{0}}\left( 1-\sin \alpha
\right) ^{2}\left\{ \sin \left( 2\theta +2\sigma \right) \times \dfrac{%
\delta ^{\left( 2\right) }}{r_{0}}+\cos \left( 2\theta +2\sigma \right)
\times \dfrac{\delta ^{\left( 3\right) }}{r_{0}}\right\} 
\end{eqnarray*}

Each of these terms, except $K_{0},$ has a specific frequency. These terms
can be measured and distinguished from each other.

The Lense-Thirring effect due to the angular momentum of the Earth appears
in the terms $K_{\sigma }$ and $K_{2\theta \pm \sigma }$ while the possible
existence of a preferred frame appears in $K_{\theta \pm \sigma }$ which
depends on the components of $\alpha _{1}\overrightarrow{w}.$

The signal due to the Lense-Thirring effect is associated with the signal
due to $\delta ^{\left( 1\right) }.$ The two signals display the same order
of magnitude when $\delta ^{\left( 1\right) }\sim 1 \, {\rm nm}.$ Today, it
seems impossible to achieve such a precision, this is the reason why $\delta
^{\left( 1\right) }$ should be calculated from the Fourier analysis of the
signal itself, altogether with the angular momentum of the Earth, $%
\overrightarrow{J}$ and the velocity $\alpha _{1}\overrightarrow{w}.$

If the sensitivity to measure the Lense-Thirring effect with an accuracy of
20\% is achieved, it should be possible to know $\alpha _{1}\overrightarrow{w%
}$ with an accuracy better than $10^{-7}.$ Considering that $\overrightarrow{%
w}$ is the velocity of the rest frame of the Universe ($\left\| 
\overrightarrow{w}\right\| \sim 10^{-3})$ it would give precision on $\alpha
_{1}$ of order of $10^{-4}.$

The interest of the spin is obvious. If $\sigma =cte$ (no spin) the signal
is the sum of two periodic signals with frequency $\nu _{O}$ and $2\nu _{O}$
where $\nu _{O}$ is the orbital frequency of the satellite~; therefore one
ASU gives two informations (two functions of the time). When the satellite
spins, we get 9 functions of the time $t.$ The information is much more
important in this case.

\section{Conclusion}

In this paper we have sketched a method to take into account the residual
gravitation in a nearly free falling satellite, namely the tidal and higher
order effects. We have shown that these effects are not negligible in highly
accurate experiments.

We have shown that many perturbations must be considered if one wants to
observe the Lense-Thirring effect and we have exhibited the various terms
that one needs to calculate in order to obtain the full signal.

Compared with GPB, the principle of the measure is not the same, the
difficulties are quite different but the job is not easier. For instance,
considering the quantities $K_{\sigma }$ or $K_{2\theta \pm \sigma }$ above,
one can check that $\delta ^{\left( 1\right) }$\ must remain smaller than $2%
 \, {\rm nm}$\ for the corresponding signal to remain smaller than the
Lense-Thirring one. It does not seem that such a precision can be controlled
in the construction of the experimental device itself. It is therefore
necessary to measure $\delta ^{\left( 1\right) }$ with such an accuracy.

In the problem that we have considered, there are 9 unknown parameters~: i)
the three component of $\overrightarrow{J},$ ii) the three components of $%
\alpha _{1}\overrightarrow{w}$ and iii) the three components of $%
\overrightarrow{\delta }.$ On the other hand, the experimental set-up
displays 9 periodic functions but the distribution of the unknown parameters
among the 9 functions happen in such way that the four parameters $%
\overrightarrow{J}$ and $\delta ^{(1)}$ are present in the 3 functions $%
K_{\sigma }$ or $K_{2\theta \pm \sigma }$ and the three parameters $\alpha
_{1}\overrightarrow{w}$ in the two functions $K_{\theta \pm \sigma }$. Only
the two parameters $\delta ^{\left( 2\right) }$ and $\delta ^{\left(
3\right) }$ are over determined by the four functions $K_{2\theta },$ $%
K_{2\sigma }$ and $K_{2\theta \pm 2\sigma }.$

\ 

Let us assume that $\theta $ and $\sigma $ are known function of the time
(frequency and phase). This implies that in the geometric scheme that we
have explored, one can determine 18 unknown parameters. Therefore $%
\overrightarrow{J},$ $\overrightarrow{\delta }$ and $\alpha _{1}%
\overrightarrow{w}$ can be known and the Lense-Thirring effect can be
observed with an accuracy of a few tens percent. The same sensitivity on the
phase difference of matter waves in the interferometers yields an accuracy
of $10^{-7}$ on $\alpha _{1}\overrightarrow{w}$ which would increase our
knowledge on $\alpha _{1}$ by one order of magnitude. This optimistic
conclusion must be tempered with the remark that only the shift has been
considered here while several other geometrical perturbations play their
role. Moreover a crucial point is the knowledge of the phase of the various
periodic functions $K$. The geometric scheme fails to describe the change of
the phase of the atomic wave when it goes through the laser beam and we
believe that the preceding conclusion holds only in the case where the
change of the phase along the two paths differs by a constant.

As a conclusion, we put forward that only a more powerful model can answer
the question of the theoretical feasibility. This model should take into
account all the gravitational perturbations that we have outlined here and
it should consider the interaction between laser fields and matter waves in
more a realistic manner.

\appendix

\section{The gravitational phase shifts}

Let us assume that space time is quasi Minkowskian. Therefore, the metric
is: $ds^{2}=\left( \eta _{\alpha \beta }+h_{\alpha \beta }\right)
\,dx^{\alpha }dx^{\beta }$ with \ \ $\left| h_{\alpha \beta }\right| <<1$.

Let us consider, at the eikonal approximation, a wave which propagates from
the point $A$ to a point $B.$ The phase at point $A$ is known. It is $%
\varphi _{A}(t)=\omega t$ where $\omega $ is a constant and $t=x^{0}/c$ the
time. The phase, $\varphi _{B}(t),$ at point $B$ is the amount of the
unperturbed phase, $\overset{{\rm o}}{\varphi }_{B}(t),$ and the
perturbation $\delta \varphi _{B}$ due to the term $h_{\alpha \beta }$ in
the metric.

Two different cases are relevant for the problem that we study

\begin{enumerate}
\item Point $A$ is a far away, fixed, star. It is the source of a light
wave. From the knowledge of $\varphi _{B}(t)$ at every point $B$ one can
deduce the apparent direction of the star. Then it becomes possible to
determine the vector $-L^{\alpha }$ which points towards $A.$

\item Point $A$ is the source of an atomic wave which enters a matter wave
interferometer. At point $B$ interferences are observed on a detector
(figure \ref{figasu}-$b$). The phases $\varphi _{B1}(t)$ and $\varphi
_{B2}(t),$ of the waves which interfere at point $B$ depend on the path, $%
(1) $ or $(2)$, that each wave has followed. The response of the detector at
point $B$ depends on the difference $\varphi _{B1}(t)-\varphi _{B2}(t).$
Therefore one can obtain the response of the interferometer once the phases $%
\varphi _{B1}(t)$ and $\varphi _{B2}(t)$ are known.
\end{enumerate}

\ 

In order to calculate the phase $\varphi _{B}(t)$ at point $B$ and time $t$
we use the method developed in \cite{LetT}. We summarize briefly the method
for particles of mass $m$ (for the light $m=0).$

First we neglect the perturbation and we consider a point $M$ which moves at
the group velocity, $v_{g}^{k},$ and arrives at point $B$ at time $t.$ The
worldline of $M$ is $x^{k}=x^{k}(t^{\prime })$ with $v_{g}^{k}=dx^{k}/dt^{\prime }$
and $x^{k}(t)=x_{B}^{k}.$ The point $M$ has left $A$ at time $t_{A}$ such as 
$x^{k}(t_{A})=x_{A}^{k}.$ The time $t_{A}$ is a function of $t.$

Now we define $\Psi =h_{00}+2h_{0k}v_{g}^{k}+h_{kj}v_{g}^{k}v_{g}^{j}$ where 
$\Psi $ is calculated at point $M.$ Therefore $\Psi $ is a function of $%
t^{\prime }.$

One can prove that the perturbation $\delta \varphi _{B}$ is 
\begin{equation*}
\delta \varphi _{B}=\dfrac{\omega }{2}\int_{t_{A}}^{t}\Psi (t^{\prime
})\,dt^{\prime }
\end{equation*}
where $\hbar \, \omega$ is the energy of the particle.

\section{Deflection of the light due to the quadrupolar terms of the Earth}

The Newtonian potential to be considered is 
\begin{equation*}
U=\dfrac{M_{\oplus }}{r}\left( 1-J_{2}\left( \dfrac{R_{0}}{r}\right)
^{2}P_{2}\right)
\end{equation*}%
where $P_{2}=\dfrac{1}{2}\left( 3\cos ^{2}\theta -1\right) $. The unitary
vector $\overrightarrow{k}$ defines the axi-symetry axis and $\cos \theta =%
\dfrac{\vec{k}\cdot \vec{r}}{r}.$ The quadrupole contribution is due to the
term $-J_{2}\dfrac{M_{\oplus }}{r}\left( \dfrac{R_{0}}{r}\right) ^{2}P_{2}.$

\begin{equation*}
\delta \varphi =-\dfrac{\omega M_{\oplus }}{4}J_{2}R_{0}^{\;2}\int_{-\infty
}^{0}\left( 3\dfrac{\left( \overrightarrow{r}^{\prime }\cdot \overrightarrow{%
k}\right) ^{2}}{r^{\prime 5}}-\dfrac{1}{r^{\prime 3}}\right) \,ds
\end{equation*}%
where $\vec{x}_{M}(t^{\prime })=\overrightarrow{r}^{\prime }=\vec{x}_{B}+s%
\vec{\ell}.$ Here $s$ is $t^{\prime }-t,$ it varies from $s_{A}=-\infty $
(when $A$ is far away) to $s_{B}=0$ at point $B.$

We calculate the gradient of $\delta \varphi $ at point $B$ the coordinates
of which are $\overrightarrow{r}=\left\{ x_{B}^{\;k}\right\} :$%
\begin{eqnarray*}
\overrightarrow{\nabla }\delta \varphi &=&-\dfrac{\omega M_{\oplus }}{4}%
J_{2}R_{0}^{\;2}\,\vec{Q}{\rm  \ with} \\
\vec{Q} &=&6\vec{k}\int_{-\infty }^{0}\dfrac{\left( \overrightarrow{r}%
^{\prime }\cdot \overrightarrow{k}\right) }{r^{\prime 5}}ds-15\,%
\overrightarrow{r}\int_{-\infty }^{0}\dfrac{\left( \overrightarrow{r}%
^{\prime }\cdot \overrightarrow{k}\right) ^{2}ds}{r^{\prime 7}}+3%
\overrightarrow{r}\int_{-\infty }^{0}\dfrac{ds}{r^{\prime 5}}
\end{eqnarray*}%
Each of these integral can be exactly computed. The deviation of the light
ray is $\Delta \vec{\ell}=\dfrac{1}{\omega }\overrightarrow{\nabla }\delta
\varphi $. We can eliminate some components parallel to $\vec{\ell}$ which
would in any case disappear in the normalization procedure. One finds

\begin{eqnarray*}
\delta \vec{\ell} &=&-\dfrac{M_{\oplus }}{4r}J_{2}\left( \dfrac{R_{0}}{r}%
\right) ^{2}\dfrac{a\overrightarrow{k}+b\overrightarrow{n}}{\left( 1-%
\overrightarrow{n}\cdot \overrightarrow{\ell }\right) ^{3}}{\rm  with} \\
a &=&2\left( \overrightarrow{n}\cdot \overrightarrow{k}\right) \left( 
\overrightarrow{n}\cdot \overrightarrow{\ell }\right) ^{2} \\
&&-6\left( \overrightarrow{n}\cdot \overrightarrow{k}\right) \left( 
\overrightarrow{n}\cdot \overrightarrow{\ell }\right) +2\left( 
\overrightarrow{\ell }\cdot \overrightarrow{k}\right) \left( \overrightarrow{%
n}\cdot \overrightarrow{\ell }\right) \\
&&+4\left( \overrightarrow{n}\cdot \overrightarrow{k}\right) -2\left( 
\overrightarrow{\ell }\cdot \overrightarrow{k}\right) \\
b &=&-3\left( \overrightarrow{n}\cdot \overrightarrow{k}\right) ^{2}\left( 
\overrightarrow{n}\cdot \overrightarrow{\ell }\right) ^{2}+9\left( 
\overrightarrow{n}\cdot \overrightarrow{k}\right) ^{2}\left( \overrightarrow{%
n}\cdot \overrightarrow{\ell }\right) \\
&&-2\left( \overrightarrow{n}\cdot \overrightarrow{k}\right) \left( 
\overrightarrow{\ell }\cdot \overrightarrow{k}\right) \left( \overrightarrow{%
n}\cdot \overrightarrow{\ell }\right) +\left( \overrightarrow{n}\cdot 
\overrightarrow{\ell }\right) ^{2}-8\left( \overrightarrow{n}\cdot 
\overrightarrow{k}\right) ^{2} \\
&&-2\left( \overrightarrow{\ell }\cdot \overrightarrow{k}\right)
^{2}+6\left( \overrightarrow{n}\cdot \overrightarrow{k}\right) \left( 
\overrightarrow{\ell }\cdot \overrightarrow{k}\right) -3\left( 
\overrightarrow{n}\cdot \overrightarrow{\ell }\right) +2
\end{eqnarray*}
where $r=\sqrt{\vec{r}^{2}}$ and $\vec{n}=\vec{r}/r.$

When $\vec{\ell}$ is orthogonal to the orbit, the maximum of $\left\| \delta 
\vec{\ell}\right\| $ is $\dfrac{M_{\oplus }}{r}J_{2}\left( \dfrac{R_{0}}{r}%
\right) ^{2}\sim 2\,10^{-3}O_{2}$ on a polar orbit of radius $r\sim 7000%
 \, {\rm km}.$

%\section*{References}

\bibliography{angonin}

\def\JOB{J. Opt. B} \def\aap{Astron. Astrophys.} \def\JMP{J. Math. Phys.}
  \def\CJP{Can. J. Phys.} \def\CQG{Class. Quantum Grav.} \def\PRL{Phys. Rev.
  Lett.} \def\PRA{Phys. Rev. A} \def\PRD{Phys. Rev. D} \def\NewA{New Astron.}
\begin{thebibliography}{15}
\expandafter\ifx\csname natexlab\endcsname\relax\def\natexlab#1{#1}\fi
\expandafter\ifx\csname bibnamefont\endcsname\relax
  \def\bibnamefont#1{#1}\fi
\expandafter\ifx\csname bibfnamefont\endcsname\relax
  \def\bibfnamefont#1{#1}\fi
\expandafter\ifx\csname citenamefont\endcsname\relax
  \def\citenamefont#1{#1}\fi
\expandafter\ifx\csname url\endcsname\relax
  \def\url#1{\texttt{#1}}\fi
\expandafter\ifx\csname urlprefix\endcsname\relax\def\urlprefix{URL }\fi
\providecommand{\bibinfo}[2]{#2}
\providecommand{\eprint}[2][]{\url{#2}}

\bibitem[{\citenamefont{{Rasel \textit{et al.}}}(2000)}]{Raselhyper}
\bibinfo{author}{\bibfnamefont{E.}~\bibnamefont{{Rasel \textit{et al.}}}},
  \emph{\bibinfo{title}{ESA Assessment Study Report}}
  (\bibinfo{publisher}{ESA-SCI}, \bibinfo{year}{2000}).

\bibitem[{\citenamefont{Jentsch et~al.}(2003)\citenamefont{Jentsch, Muellerand,
  Chelkowski, Rasel, and Ertmer}}]{Rasel1}
\bibinfo{author}{\bibfnamefont{C.}~\bibnamefont{Jentsch}},
  \bibinfo{author}{\bibfnamefont{T.}~\bibnamefont{Muellerand}},
  \bibinfo{author}{\bibfnamefont{S.}~\bibnamefont{Chelkowski}},
  \bibinfo{author}{\bibfnamefont{E.}~\bibnamefont{Rasel}}, \bibnamefont{and}
  \bibinfo{author}{\bibfnamefont{W.}~\bibnamefont{Ertmer}},
  \bibinfo{journal}{Verhanal DPG (VI)} \textbf{\bibinfo{volume}{38}},
  \bibinfo{pages}{167} (\bibinfo{year}{2003}).

\bibitem[{\citenamefont{Oberthaler et~al.}(1996)\citenamefont{Oberthaler,
  Bernet, Rasel, Schmiedmayer, and Zeilinger}}]{Rasel2}
\bibinfo{author}{\bibfnamefont{M.}~\bibnamefont{Oberthaler}},
  \bibinfo{author}{\bibfnamefont{S.}~\bibnamefont{Bernet}},
  \bibinfo{author}{\bibfnamefont{E.}~\bibnamefont{Rasel}},
  \bibinfo{author}{\bibfnamefont{J.}~\bibnamefont{Schmiedmayer}},
  \bibnamefont{and}
  \bibinfo{author}{\bibfnamefont{A.}~\bibnamefont{Zeilinger}},
  \bibinfo{journal}{\PRA} \textbf{\bibinfo{volume}{54}}, \bibinfo{pages}{3165}
  (\bibinfo{year}{1996}).

\bibitem[{\citenamefont{Gustavson et~al.}(2000)\citenamefont{Gustavson,
  Landragin, and Kasevich}}]{belarno}
\bibinfo{author}{\bibfnamefont{T.}~\bibnamefont{Gustavson}},
  \bibinfo{author}{\bibfnamefont{A.}~\bibnamefont{Landragin}},
  \bibnamefont{and} \bibinfo{author}{\bibfnamefont{M.}~\bibnamefont{Kasevich}},
  \bibinfo{journal}{\CQG} \textbf{\bibinfo{volume}{17}}, \bibinfo{pages}{2385}
  (\bibinfo{year}{2000}).

\bibitem[{\citenamefont{{Le Coq} et~al.}(2001)\citenamefont{{Le Coq},
  Thywissen, Rangwala, Gerbier, Richard, Delannoy, Bouyer, and
  Aspect}}]{Bouyer1}
\bibinfo{author}{\bibfnamefont{Y.}~\bibnamefont{{Le Coq}}},
  \bibinfo{author}{\bibfnamefont{J.}~\bibnamefont{Thywissen}},
  \bibinfo{author}{\bibfnamefont{S.}~\bibnamefont{Rangwala}},
  \bibinfo{author}{\bibfnamefont{F.}~\bibnamefont{Gerbier}},
  \bibinfo{author}{\bibfnamefont{R.}~\bibnamefont{Richard}},
  \bibinfo{author}{\bibfnamefont{G.}~\bibnamefont{Delannoy}},
  \bibinfo{author}{\bibfnamefont{P.}~\bibnamefont{Bouyer}}, \bibnamefont{and}
  \bibinfo{author}{\bibfnamefont{A.}~\bibnamefont{Aspect}},
  \bibinfo{journal}{\PRL} \textbf{\bibinfo{volume}{87}},
  \bibinfo{pages}{170403} (\bibinfo{year}{2001}).

\bibitem[{\citenamefont{Snadden et~al.}(1998)\citenamefont{Snadden, McGuirk,
  Bouyer, Haritos, and Kasevich}}]{Bouyer2}
\bibinfo{author}{\bibfnamefont{M.}~\bibnamefont{Snadden}},
  \bibinfo{author}{\bibfnamefont{J.}~\bibnamefont{McGuirk}},
  \bibinfo{author}{\bibfnamefont{P.}~\bibnamefont{Bouyer}},
  \bibinfo{author}{\bibfnamefont{K.}~\bibnamefont{Haritos}}, \bibnamefont{and}
  \bibinfo{author}{\bibfnamefont{M.}~\bibnamefont{Kasevich}},
  \bibinfo{journal}{\PRL} \textbf{\bibinfo{volume}{81}}, \bibinfo{pages}{971}
  (\bibinfo{year}{1998}).

\bibitem[{\citenamefont{Ni and Zimmermann}(1978)}]{LiZ}
\bibinfo{author}{\bibfnamefont{W.-T.} \bibnamefont{Ni}} \bibnamefont{and}
  \bibinfo{author}{\bibfnamefont{M.}~\bibnamefont{Zimmermann}},
  \bibinfo{journal}{\PRD} \textbf{\bibinfo{volume}{17}}, \bibinfo{pages}{1473}
  (\bibinfo{year}{1978}).

\bibitem[{\citenamefont{Li and Ni}(1979)}]{LiNi}
\bibinfo{author}{\bibfnamefont{W.-Q.} \bibnamefont{Li}} \bibnamefont{and}
  \bibinfo{author}{\bibfnamefont{W.-T.} \bibnamefont{Ni}},
  \bibinfo{journal}{\JMP} \textbf{\bibinfo{volume}{20}}, \bibinfo{pages}{1473}
  (\bibinfo{year}{1979}).

\bibitem[{\citenamefont{Antoine and Bord\'e}(2003)}]{Antoine}
\bibinfo{author}{\bibfnamefont{C.}~\bibnamefont{Antoine}} \bibnamefont{and}
  \bibinfo{author}{\bibfnamefont{C.}~\bibnamefont{Bord\'e}},
  \bibinfo{journal}{\JOB} \textbf{\bibinfo{volume}{5}}, \bibinfo{pages}{S199}
  (\bibinfo{year}{2003}).

\bibitem[{\citenamefont{Will}(1981)}]{will}
\bibinfo{author}{\bibfnamefont{C.}~\bibnamefont{Will}},
  \emph{\bibinfo{title}{Theory and experiment in gravitational physics}}
  (\bibinfo{publisher}{Cambridge University Press}, \bibinfo{year}{1981}).

\bibitem[{\citenamefont{Marchal}(1996)}]{march}
\bibinfo{author}{\bibfnamefont{C.}~\bibnamefont{Marchal}},
  \bibinfo{journal}{Bulletin du Mus\'{e}um National d'Histoire Naturelle 4\`eme
  s\'erie section C} \textbf{\bibinfo{volume}{18}}, \bibinfo{pages}{517}
  (\bibinfo{year}{1996}).

\bibitem[{\citenamefont{Linet and Tourrenc}(1976)}]{LetT}
\bibinfo{author}{\bibfnamefont{B.}~\bibnamefont{Linet}} \bibnamefont{and}
  \bibinfo{author}{\bibfnamefont{P.}~\bibnamefont{Tourrenc}},
  \bibinfo{journal}{\CJP} \textbf{\bibinfo{volume}{54}}, \bibinfo{pages}{1129}
  (\bibinfo{year}{1976}).

\bibitem[{\citenamefont{Ib\'a\~nez}(1983)}]{Ibanez}
\bibinfo{author}{\bibfnamefont{J.}~\bibnamefont{Ib\'a\~nez}},
  \bibinfo{journal}{\aap} \textbf{\bibinfo{volume}{124}}, \bibinfo{pages}{175}
  (\bibinfo{year}{1983}).

\bibitem[{\citenamefont{Touboul and Rodrigues}(2001)}]{Touboul}
\bibinfo{author}{\bibfnamefont{P.}~\bibnamefont{Touboul}} \bibnamefont{and}
  \bibinfo{author}{\bibfnamefont{M.}~\bibnamefont{Rodrigues}},
  \bibinfo{journal}{\CQG} \textbf{\bibinfo{volume}{18}}, \bibinfo{pages}{2487}
  (\bibinfo{year}{2001}).

\bibitem[{\citenamefont{Nobili et~al.}(2003)\citenamefont{Nobili, Bramanti, and
  {Comandi \textit{et al.}}}}]{Nobili}
\bibinfo{author}{\bibfnamefont{A.}~\bibnamefont{Nobili}},
  \bibinfo{author}{\bibfnamefont{D.}~\bibnamefont{Bramanti}}, \bibnamefont{and}
  \bibinfo{author}{\bibfnamefont{G.}~\bibnamefont{{Comandi \textit{et al.}}}},
  \bibinfo{journal}{\NewA} \textbf{\bibinfo{volume}{8}}, \bibinfo{pages}{371}
  (\bibinfo{year}{2003}).

\end{thebibliography}

\end{document}